\documentclass[prb,superscriptaddress,aps, bibliography, twocolumn, reprint, showpacs, footnoteinbib]{revtex4-1}

\usepackage{graphicx}
\usepackage{amssymb}   
\usepackage{amsfonts}
\usepackage{amsmath}
\usepackage{dsfont}
\usepackage{natbib}
\usepackage{dcolumn}   
\usepackage{bm}        
\usepackage[mathscr]{eucal}
\usepackage[dvipsnames]{xcolor}
\usepackage[colorlinks, linkcolor=OrangeRed,citecolor=RoyalBlue,urlcolor=NavyBlue]{hyperref}
\usepackage[all]{hypcap} 
\usepackage{mathtools}
\usepackage{wasysym}

\usepackage{graphicx}
\usepackage{amsmath,amssymb,bm}
\usepackage{enumerate}
\usepackage{braket}        

\begin{document}

\title{Algebraic Many-Body Localization and its implications on information propagation}
\author{Giuseppe De Tomasi}
\affiliation{Max-Planck-Institut f\"ur Physik komplexer Systeme, N\"othnitzer Stra{\ss}e 38,  01187-Dresden, Germany}
\begin{abstract}
We probe the existence of a many-body localized phase (MBL-phase) in a spinless fermionic Hubbard chain with algebraically localized single-particle states, 
by investigating both static and dynamical properties of the system.
This MBL-phase can be characterized by an extensive number of integrals of motion which develop algebraically decaying tails, unlike the case of exponentially localized
single-particle states.
We focus on the implications for the quantum information propagation through the system.  
We provide evidence that the bipartite entanglement entropy after a quantum quench has an unbounded algebraic growth in time, while the quantum Fisher information grows 
logarithmically. 
\end{abstract}
\maketitle

Anderson localization is a wave phenomenon in which transport in non-interacting fermionic
systems can be suppressed due to the presence of a quenched disorder potential~\cite{Anderson58,0034-4885-56-12-001}, which localizes the
single-particle eigenstates exponentially in space ~\cite{Goldshtein1977}.
The question whether ergodicity is restored once the localized single-particle eigenstates interact with each other has been 
debated for several decades~\cite{0022-3719-15-36-018,PhysRevB.21.2366,PhysRevLett.95.206603,PhysRevLett.78.2803,1990TMP85.1223B}, culminating in the discovery of a metal-insulator transition (MIT) at finite energy-density~\cite{Basko06}. 
This MIT separates a delocalized phase, which is believed to be thermal, from an ergodicity breaking phase known as the many-body localized (MBL-) phase~\cite{Pal10, Basko06, doi:10.1146/annurev-conmatphys-031214-014726, 2018arXiv180411065A,ALET2018}. 
The MBL-phase describes a perfect many-body insulator, whose eigenstates are adiabatically connected 
to the non-interacting localized ones, implying a full description in terms of an extensive number of 
quasi-local integrals of motion (LIOMs)~\cite{Chandran15,Ros15,Huse14, Aba13,PhysRevB.97.094206, doi:10.1002/andp.201600322, Rade16}. The existence of the LIOMs emphasizes an emergent weak form of integrability~\cite{Huse14,Imbrie2016,Bala13}.
As a result, an MBL-phase can be described with arbitrary precision by an integrable localized quantum system, often called $l$-bit Hamiltonian~\cite{Huse14}.
Although both the Anderson localized phase and the MBL-phase have similar static properties, their dynamics are intrinsically different; interactions weakly couple 
the LIOMs inducing a dephasing mechanism which leads to a logarithmically slow entanglement propagation~\cite{Bar12, Proz08, Pep13, Ze18,PhysRevB.96.174201,2018arXiv180801250S}.

The existence of the MBL-phase has been confirmed in several numerical works~\cite{Luitz15, Kill14, Pal10, 0295-5075-101-3-37003,Serb15, Hell17, PhysRevB.91.094202, 2018arXiv181101925H} as well as experimentally 
using quantum simulators, such as ultracold atoms in optical lattices~\cite{schreiber15,Choi16,Bordia16,Lus17} and trapped ions~\cite{Smith2016}.  
Moreover, systems where the single-particle spectrum is fully (or partially) delocalized have also been found to show MBL~\cite{PhysRevA.95.021601, 2017arXiv170708845S, PhysRevB.94.201116,PhysRevLett.115.230401}.
Thus it is  not necessary to have exponentially localized single-particle states \textit{a priori} for MBL to exist.
How the $l$-bit picture changes in an MBL-phase where the single-particle eigenstates are not exponentially localized and its implications for information propagation, has not been explored so far.

The aim of this work is to investigate the possible existence of an MBL-phase in the case where the single-particle eigenstates are algebraically localized~\cite{PhysRevLett.120.110602,2018arXiv181001492N} ($\psi_E (x)\sim 1/x^{\alpha}$). 
We numerically show that an \textit{algebraic} MBL-phase is indeed possible. In such a phase the integrals of motion develop algebraically decaying tails.
This is different from the case with exponentially localized single-particle states, where the corresponding MBL-phase has exponentially decaying LIOMs.  

Our study focus on quantum information propagation in this algebraic MBL-phase. 
In particular, we study the dynamics after a global quench and show that this MBL-phase is characterized by an unbounded algebraic growth of entanglement.  
We also investigate quantum information transport through the quantum Fisher information~\cite{Hauke2016}, and find that its propagation is logarithmically slow, 
in agreement with a recent experiment in trapped ions, where a long-range disordered Ising model was studied~\cite{Smith2016}. 

The rest of the paper is organized as follows. In Sec.~\ref{sec:sec_1}, we introduce the model and discuss the non-interacting limit 
in which its single-particle eigenstates are algebraically localized. In Sec.~\ref{sec:sec_21}, we present numerical evidence of the existence of
an interacting localized phase, studying static diagnostics, such as energy level spacing statics, entanglement properties and fluctuations 
of local observables. In Sec.~\ref{sec:sec_22}, to provide further evidence, non-equilibrium dynamical properties of the model are inspected. In particular, we 
center our attention on the imbalance~\cite{doi:10.1002/andp.201600350}, which has been used as a dynamical indicator for ergodicity breaking in several recent experiments~~\cite{schreiber15,Choi16,Bordia16,Lus17}. 
In Sec.~\ref{sec:sec_3}, we study the quantum mutual information~\cite{RevModPhys.80.517} and the long-time density-density correlator~\cite{Chandran15,Bera17}, and show that the LIOMs can have algebraically decaying tails. 
While the quantum mutual information gives us information about correlation functions, the long-time limit of the density-density correlator lets us access the LIOMs directly.
In Sec.~\ref{sec:sec_4}, we investigate the information propagation in this localized phase, focusing on the dynamics of bipartite entanglement entropy and quantum Fisher information 
following a quantum quench starting from a charge-density wave. 

\section{Model}
\label{sec:sec_1}
We study the spinless fermionic Hubbard Hamiltonian in one-dimension
 \begin{equation} 
 \hat{H} =\sum_{x\ne y } (J_{x,y} \hat{c}^\dagger_{x} \hat{c}_{y} + h.c.)  + \sum_{x}  h_x \hat{n}_x + V\sum_{x}  \hat{n}_x \hat{n}_{x+1},
\label{eq:Ham}
\end{equation}
where $\hat{c}_x^\dagger~(\hat{c}_x)$ is the fermionic creation (annihilation) operator at
site $x$ and $\hat{n}_x = \hat{c}_x^\dagger \hat{c}_x $ is the particle-density operator. 
The constants $\{J_{x,y}\}$ denote the long-range hopping between sites $x$ and $y$, and are defined as $J_{x,y} = J_{y,x} = \mu_{x,y}/(1+|x-y|^{2\alpha})^{1/2}$. 
While $\alpha$ sets the power-law decay exponent of the hopping and $\{\mu_{x,y}\}$ are independent uniformly distributed random  variables  between $[-1,1]$. 
$\{h_x\}$ are random fields distributed between $[-W,W]$ and $V$ is the interaction strength. $L$ is the length of the chain, $N=L/2$ the number of fermions and boundary conditions are taken to be open.
The average over disorder will be indicated with an overline over the quantity taken into consideration (e.g. $\overline{A}$). 

The non-interacting limit of $\hat{H}$ is known as the power-law random banded matrix model ($V=0$), and it is known to have a MIT as a function of $\alpha$~\cite{Mirlin00,Mirlin96,PhysRevLett.64.547,Mirlin93,0295-5075-9-1-015}.
For $V=0$ its single-particle eigenstates $\{\psi_E (x)\}$ for $\alpha<1$ are delocalized, 
$\sum_x |\psi_E(x)|^{2q}\sim L^{(1-q)}$ ($q\ge 1$), and for $\alpha > 1$  $\{\psi_E (x)\}$ are algebraically localized $\psi_E (x)\sim 1/x^{\alpha}$. In the limit $\alpha \rightarrow \infty$ 
the single-particle eigenstates of $\hat{H}$ are exponentially localized and one 
recovers the paradigmatic model having an MBL-transition~\cite{Pal10,Bera15,Proz08,Serb15,Bera16,Ser14,Aban16,Vass15,Luitz15,1367-2630-19-11-113021,2018arXiv181101925H}.

\begin{figure}
	\includegraphics[width=1.\columnwidth]{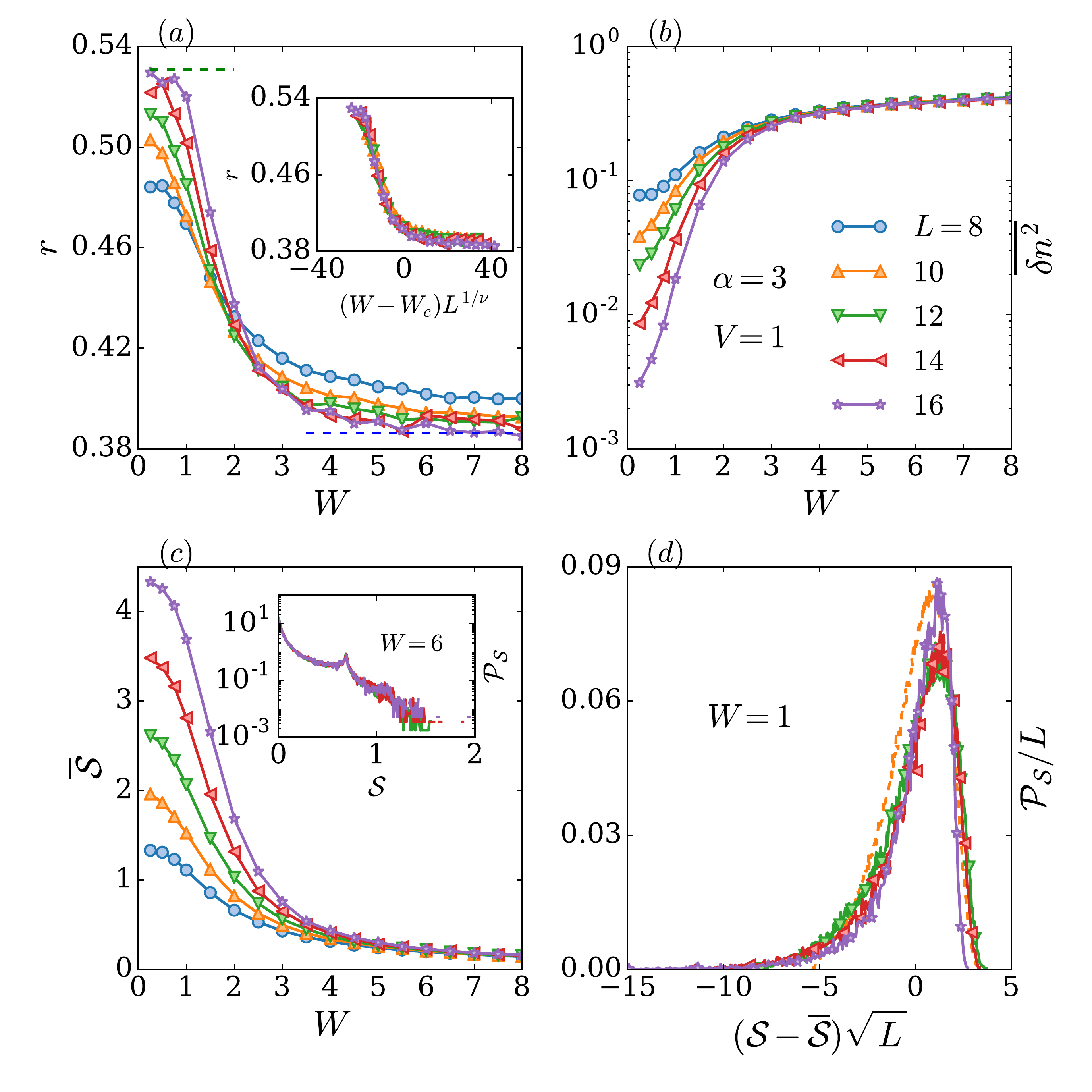}
	\caption{(a): Level statistics as a function of disorder strength $W$ and fixed $V$ for several system sizes $L$. 
	 The upper dashed line is $r_{\text{GOE}} \approx 0.5307$, the value of $r$ in case of Wigner-Dyson level statistic. The lower dashed line is $r_{\text{Poisson}} =  2\log{2}-1 \approx 0.3863$, the value of $r$ in the case of
	 Poisson level statistic. Its inset shows $r$ level statistics parameter as a function of $(W-W_c)L^{\mu}$ with $W_c=3.1$ and $\nu=1.3$.
	(b): Fluctuations of the expectation value of the density operator $\delta n$ as a function of $W$ at fixed $V$ and several $L$. 
	(c): Disorder averaged bipartite entanglement entropy $\overline{\mathcal{S}}$ as a function of $W$ for fixed $V$ and several $L$. Its inset  shows the probability distribution $\mathcal{P}_\mathcal{S}$
	of $\mathcal{S}$ for several $L$ in the localized phase $W=6$. (d) Probability distribution $\mathcal{P}_\mathcal{S}$
	of $\mathcal{S}$ for several $L$ in the delocalized phase $W=1$. The curves have been rescaled to underline $\text{Var}[\mathcal{S}]\sim 1/L$ and $\max_\mathcal{S} \mathcal{P}_\mathcal{S}\sim L$. } 
	\label{fig:Fig1}
\end{figure}

In this work we focus on the case in which the single-particle eigenstates of $\hat{H}$ (Eq.~\ref{eq:Ham}) are algebraically localized, hence we choose 
$\alpha>1$. 

\section{Spectral $\&$ Eigenstate analysis}\label{sec:sec_2}
\subsection{Static Properties}\label{sec:sec_21}
In this section we inspect static properties of $\hat{H}$ with the aim to show the existence of an MBL-phase.

The resistance to have crossing of energy levels (level repulsion) is a well known property of ergodic systems~\cite{Haake:2006:QSC:1214825,Luca16,2018arXiv180506472D}. 
The level spacing statistics of an ergodic system is expected to be the same as that of a random matrix belonging to the same
symmetry class~\cite{Haake:2006:QSC:1214825}. Thus the level spacing distribution in an ergodic phase of $\hat{H}$ should be the Wigner-Dyson distribution~\cite{Haake:2006:QSC:1214825,Luca16}. 
On the other hand, in a localized phase, due to the existence of LIOMs and thus an emergent integrability, energy levels tend to cross each other and the expected 
probability distribution for the level spacing is the Poisson distribution~\cite{Haake:2006:QSC:1214825,Luca16}. A powerful method to distinguish 
these two cases is the so called $r$ level statistics~\cite{PhysRevB.75.155111, Luitz15,PhysRevLett.110.084101}. Defining the $r$ level spacing parameter by $r^{(n)} = \min{(\delta^{(n)},\delta^{(n+1)})}/\max{(\delta^{(n)},\delta^{(n+1)})}$, where 
$\delta^{(n)} = E_{n+1} - E_{n}$ is the difference between two consecutive many-body energy levels, and averaging over the eigenstate index $n$, in the ergodic phase 
$r_{\text{GOE}}\approx 0.5306$~\cite{PhysRevB.75.155111,PhysRevLett.110.084101} and in the localized one $r_{\text{Poisson}} = 2\log{2}-1$~\cite{PhysRevB.75.155111,PhysRevLett.110.084101}. Nevertheless, it is important to keep in mind the caveat that $r$ can have 
the $r_{\text{GOE}}$ value even if the system is non-ergodic, for example in the case of fractal states~\cite{Kra17}. 

Figure~\ref{fig:Fig1} (a) shows the $r$ level spacing parameter as a function of $W$, averaged over both a few eigenstates of $\hat{H}$ in the middle of the spectrum~\footnote{We took $16$ eigenvalues in the middle of the spectrum of $\hat{H}$.} and disorder configurations, for several system
sizes $L$ and interaction strength $V=1$. We set  $\alpha = 3$ to ensure that the single-particle states are algebraically localized also for relatively small system sizes. 
At weak disorder strength $W$, $r$ reaches the value $r_{\text{GOE}}$, while at stronger disorder, $r$ approaches $r_{\text{Poisson}}$, providing evidence of the existence of the 
two phases at least for the system sizes considered.  Moreover, using finite-scaling techniques we estimate the value of the transition point $W_{c}$ between the two phases, by collapsing 
the curves for several $L$ as a function of $(W-W_c)L^{1/\nu}$, as shown in the inset of Fig.~\ref{fig:Fig1} (a). We estimate $W_c \sim 3.1$ and $\nu \sim 1.3$, however due to limitation in
system size, we can not rule out the possibility of small logaritimic corrections ($W_c\sim \log{L}$)~\cite{PhysRevB.97.214205}, which will imply the non existence of a localized phase. 

A different quantity that can be studied to shed light on the existence of an MBL-phase is the fluctuation of local observables in eigenstates which belong to the same 
energy-density. We define
\begin{equation}
 \delta n^2 =  \text{Var} [ \langle E_{n} | \hat{n}_x| E_{n} \rangle ]_E,
\end{equation}
where $\langle E_{n} | \hat{n}_x| E_{n} \rangle $ is the expectation value of  $\hat{n}_x$ in an eigenstate of $\hat{H}$ and $\text{Var}[\cdot]_E$ is the variance among few eigenstates. 
The eigenstate thermalization hypothesis (ETH)~\cite{Luca16,2008Natur.452..854R, 1994PhRvE..50..888S, 1991PhRvA..43.2046D} states that in the thermal phase the expectation value  $\langle E_{n} | \hat{n}_x| E_{n} \rangle $ is a smooth function of the energy-density and $\delta n^2$ is exponentially suppressed in system size
($\delta n^2 \sim e^{-c L}$). Instead, in the localized phase ergodicity is broken and the difference $|\langle E_{n} | \hat{n}_x| E_{n} \rangle - \langle E_{n+1} | \hat{n}_x| E_{n+1} \rangle | \sim \mathcal{O}(L^{0})$~\cite{Pal10}, thus the 
fluctuation does not scale to zero with $L$ ($ \delta n^2 \sim \mathcal{O}(L^{0})$). 

Figure~\ref{fig:Fig1} (b) shows $\overline{\delta n^2}$  as a function of $W$ for several system sizes $L$. $\overline{\delta n^2}$ has been 
computed using few eigenstates of $\hat{H}$ in the middle of the spectrum, therefore in 
the thermodynamic limit ($L\rightarrow \infty$) they will converge to the same energy-density. 
For weak disorder $W$, $\overline{\delta n^2}$  decays exponentially to zero with $L$ as expected in a thermal phase.
On the other hand, at strong disorder the curves are converged with $L$ and ergodicity is broken. 
Moreover, the value of $W_c$ is consistent with the one obtained from the $r$ level statistic. 

Finally, we investigate entanglement properties of the eigenstates of $\hat{H}$, using the bipartite half-chain entanglement entropy~\cite{RevModPhys.80.517} 
\begin{equation}
\label{eq:ent}
\mathcal{S} =-\text{Tr} \hat{\rho}_{L/2} \log{\hat{\rho}_{L/2}}, 
\end{equation}
where $\hat{\rho}_{L/2}$ denotes the  reduced  density  matrix  of half of the system in an eigenstate in the middle of the spectrum. 
$\mathcal{S}$ gives us the information on how one half of the chain 
is correlated to the other. $\mathcal{S}$ has been used to describe the MBL-transition in the case where the single-particle states are exponentially localized (i.e. $\alpha\rightarrow \infty$).
In the ergodic phase, the two halves of the system are highly entangled and $\mathcal{S}$ has a volume-law scaling~\cite{Luitz15, Pal10, Kill14}, $\mathcal{S}\sim L$. In the localized phase, only near by sites are entangled 
and $\mathcal{S}$ has an area-law, implying that in a one-dimensional system $\mathcal{S}\sim \mathcal{O}(L^0)$~\cite{Luitz15, Pal10, Kill14}. 

Figure~\ref{fig:Fig1}~(c) shows the disorder averaged $\mathcal{S}$ for eigenstates in the middle of spectrum of $\hat{H}$ as a function of $W$ for several $L$. 
In agreement with previous results, $\overline{\mathcal{S}}$ shows two different scalings with $L$ depending on the value of $W$. 
For small $W$,  $\overline{\mathcal{S}}$ has a volume-law scaling, while at large $W$, $\overline{\mathcal{S}}$ does not scale with $L$ (area-law), consistent  with the existence of a delocalized and a
localized phase respectively. Figures~\ref{fig:Fig1}~(c),~(d) show the probability distribution $\mathcal{P}_{\mathcal{S}}$ of $\mathcal{S}$ for two different values of $W$, one in the delocalized phase ($W=1$) (Fig.~\ref{fig:Fig1}~(d)) and 
another in the localized phase ($W=6$) (inset of Fig.~\ref{fig:Fig1} (c)). In the delocalized phase, the width of $\mathcal{P}_{\mathcal{S}}$ shrinks to zero in the thermodynamic limit with $\text{Var}[\mathcal{S}] \sim 1/L$ 
and $\max_{\mathcal{S}} \mathcal{P}_{\mathcal{S}} \sim  L$, converging thus to a delta function, as expected in a thermal phase~\cite{Luitz15}.  In the localized phase,  $\mathcal{P}_{\mathcal{S}}$ does not scale with $L$, since 
$\mathcal{S}$ obeys an area-law, and in addition $\mathcal{P}_{\mathcal{S}}$ develops exponential tails. 

\subsection{Dynamical Properties}\label{sec:sec_22}
In order to provide further evidence for the existence of a localized phase, we inspect dynamical properties of the model. 
A useful dynamical probe to distinguish a delocalized phase from a localized one is the imbalance, which has been used in several experiments in cold-atoms systems~\cite{schreiber15,Choi16,Bordia16,Lus17}.
This quantity is defined as~\cite{schreiber15,Choi16,Bordia16,Lus17,doi:10.1002/andp.201600350} 
\begin{equation}
 I(t) = \frac{\sum_x (-1)^x \langle \hat{n}_{x}(t) \rangle }{N}, 
\end{equation}
where $\hat{n}_{x}(t) = e^{it\hat{H}}\hat{n}_x e^{-it\hat{H}}$ with the initial state being  the charge-density wave $|\psi\rangle = \prod_{s=1}^{L/2} \hat{c}^{\dagger}_{2s} |0\rangle$, thus ensuring
$I(0)=1$. If the system is ergodic, $I(t)$ decays to zero with time, as the long-time limit of $\langle \hat{n}_{x}(t) \rangle$ depends only on the energy-density of the initial state, thus losing the memory of 
the initial state. In contrast, in the localized phase, $I(t)>0$ also in the long-time limit, thus conserving some memory of the local structure of $|\psi\rangle $. 
\begin{figure}
	\includegraphics[width=1.\columnwidth]{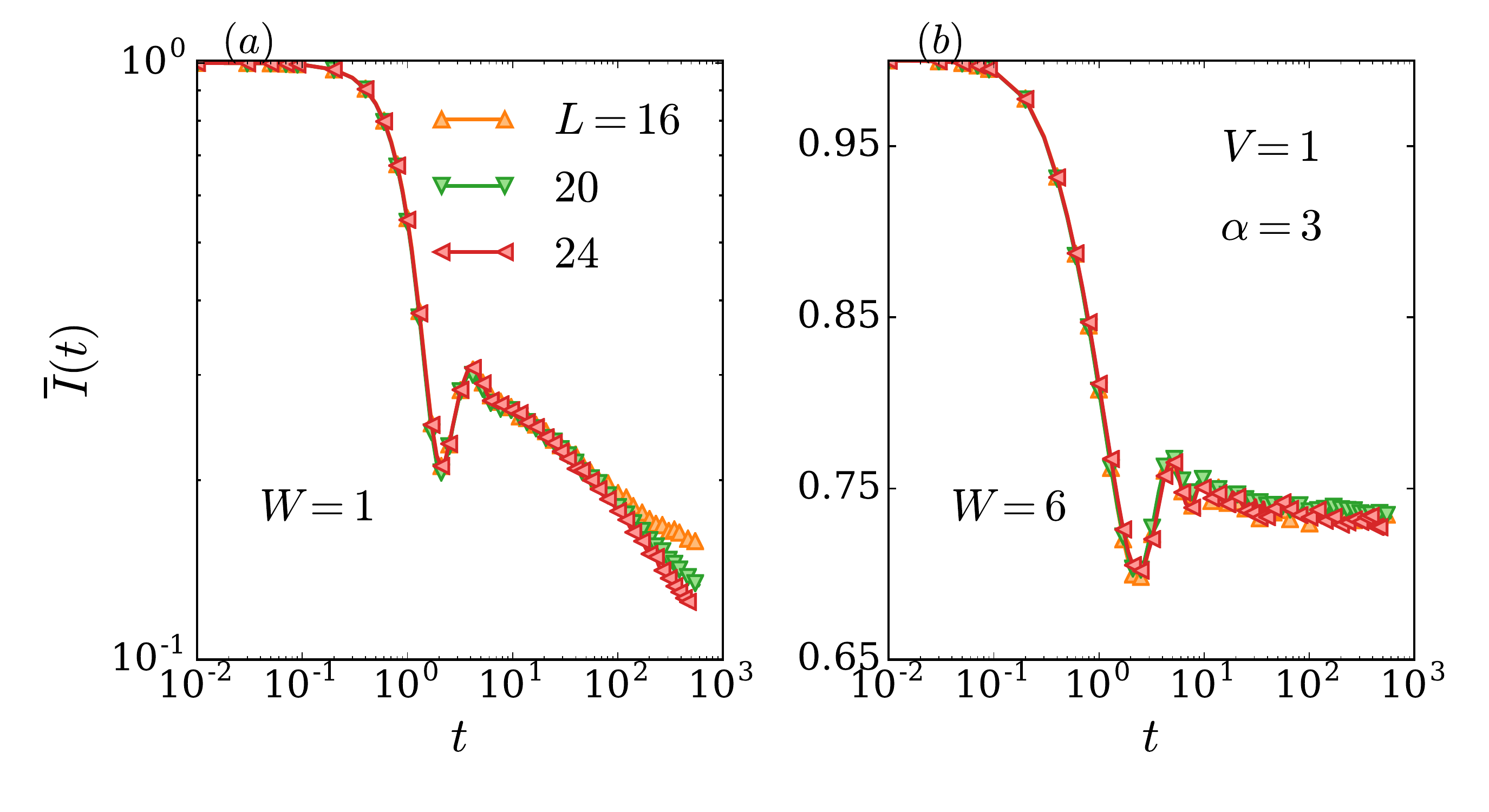}
	\caption{(a), (b): The disorder averaged imbalance $\overline{I}(t)$ for $W=1$ (delocalized phase) and $W=6$ (localized phase) respectively for several system sizes $L$ and $\alpha=3$. The initial state is the charge-density wave  
	$|\psi\rangle = \prod_{s}^{L/2} \hat{c}^{\dagger}_{2s} |0\rangle$.} 
	\label{fig:Fig2}
\end{figure}

Figures~\ref{fig:Fig2}~(a),~(b) show the imbalance averaged over disorder $\overline{I}(t)$ as a function of time for several system sizes, which has been computed using Chebyshev integration techniques~\cite{Bera17}. 
For $W=1$, $\overline{I}(t)$ shows delocalization, decaying with time, as shown in Fig.~\ref{fig:Fig2}~(a). After a fast decay at short times $(\sim \mathcal{O}(1))$, a slower decay takes place. 
For $t \lesssim 80$ the decay is consistent with a power law ($\overline{I}(t)\sim t^{-a_1}$), with $a_1< 0.5$ indicating that the transport at least within these time scales is sub-diffusive~\cite{doi:10.1002/andp.201600350}. 
However, for longer times $t \gtrsim 80$, a clear bending is visible, which could be the indication of a new time regime with a faster decaying of $I(t) \sim t^{-a_2}$ with $a_2> a_1$. 
Due to strong finite-size effects we are not able to extract the power $a_2$, since it is still $L$ dependent. For this reason, we can not rule out the possibility that diffusion  ($a_2=1/2$) could be restored at larger time scales in the thermodynamic limit~\cite{Bera17}. 
For $W=6$, $\overline{I}(t)$ shows a typical ergodicity breaking behavior saturating with time ($\lim_{t\rightarrow \infty} \overline{I}(t)>0$), and thus giving further indication of the existence of a localized phase.

\section{Algebraic Localization}\label{sec:sec_3}

Having given numerical evidence of the existence of a localized phase, we now turn to questions regarding its nature. 

The MBL-phase in the limit $\alpha \rightarrow \infty$ is characterized by the existence of the LIOMs $\{\hat{\tau}_x \}$~\footnote{Usually $\{\hat{\tau}_x\}$ are chosen to have binary spectrum (i.e. $\{0,1\}$, $\{-1/2,1/2\}$).} 
which are exponentially localized in space~\cite{Chandran15,Ros15,Huse14, Aba13,PhysRevB.97.094206, doi:10.1002/andp.201600322, Rade16}
\begin{equation}
\label{spred_int}
 \frac{1}{\mathcal{N}}\text{Tr}[\hat{\tau}_x \hat{A}_y] \sim e^{-|x-y|/\xi},
\end{equation}
where $\hat{A}_y$ is a local observable at site $y$, $\mathcal{N} =  \binom{L}{N}$ is the dimension of the Hilbert space, and $\xi$ is the localization length. 
Moreover, it is believed that the LIOMs in this case are adiabatically connected to
the integrals of motion of the non-interacting case ($V=0$) via a sequence of local unitary transformations (finite-depth unitary operations), which gives rise to an emergent integrability~\cite{Huse14,Imbrie2016,Bala13}. 
Expressing the Hamiltonian $\hat{H}$ in terms of $\{\hat{\tau}_x\}$
\begin{equation}
\label{eq:Ham_tru}
 \hat{H} = \sum_j \sum_{x_1<x_2<\cdots <x_j} b^{(j)}_{x_1,x_2,..,x_j} \hat{\tau}_{x_1}\cdots \hat{\tau}_{x_j},
\end{equation}
with $b^{(j)}_{x_1,x_2,..,x_j}\sim e^{-\min_{i,j}|x_i-x_j|/\xi}$, we can approximate $\hat{H}$ to arbitrary precision (also in the thermodynamic limit) by truncating the sum over $j$ in Eq.~\ref{eq:Ham_tru}.
It is important to note that a similar Hamiltonian representation (Eq.~\ref{eq:Ham_tru}) exists for any finite system  also in an ergodic phase. However, 
in this case the integrals of motion are spread over the chain  $\frac{1}{\mathcal{N}}\text{Tr}[\hat{\tau}_x \hat{A}_y] \sim 1/L$, and the 
sum in Eq.~\ref{eq:Ham_tru} cannot be truncated once the limit $L\rightarrow \infty$ is taken~\cite{Rade16}. 

Assuming that the adiabatic connection between non-interacting and interacting eigenstates holds also in the localized phase of $\hat{H}$ (Eq.~\ref{eq:Ham}), we can expect that $\{\hat{\tau}_x\}$ for the interacting case are algebraically localized  
($\frac{1}{\mathcal{N}}\text{Tr}[\hat{\tau}_x \hat{A}_y] \sim 1/|x-y|^{\beta}$) as in the non-interacting case. 
For $V=0$ one could take $\{\hat{\tau}_x = \hat{\eta}_x^\dagger \hat{\eta}_x\}$, 
where $\hat{\eta}_x^\dagger$ ($\hat{\eta}_x$) is the fermionic creation (annihilation) operator of the single-particle eigenstate $\psi_x$ of $\hat{H}(V=0)$ with localization center 
at site $x$ and single-particle energy $\epsilon_x$ ($\hat{H}(V=0) = \sum_x \epsilon_x \hat{\eta}^\dagger_x \hat{\eta}_x$).
For example, choosing $\hat{A}_y = \hat{n}_y$, therefore;
\begin{equation}
 \frac{1}{\mathcal{N}}\text{Tr}[ \hat{\tau}_x \hat{n}_y] = \frac{|\psi_x(y)|^2}{2} + c_L(1-|\psi_x(y)|^2) \sim \frac{1}{|x-y|^{2\alpha}},  
\end{equation}
with $c_L = \frac{1}{4} \frac{L-2}{L-1}$.

We now provide numerical evidence for the validity of our assumptions. We focus on two different quantities:
First, we consider the quantum mutual information (QMI) for two sites of the chain in eigenstates of $\hat{H}$,
which gives us information about the decay of two-point correlation functions. Second, with the aim to get insight into the integrals of motion of $\hat{H}$, we study
the long-time behavior of the density-density correlator.

The QMI for two sites of the chain is defined as
\begin{equation}
\mathcal{I}_x = \mathcal{S} ([x_0]) + \mathcal{S}( [x_0+x]  ) - \mathcal{S} ( [x_0] \cup [x_0 + x] ),
\end{equation}
where $\mathcal{S}(\mathcal{A})$ is the entanglement entropy, defined in Eq.~\ref{eq:ent}, with $\hat{\rho}_{\mathcal{A}}$ the reduced density matrix of the chain-subset $\mathcal{A}$ (e.g. $\mathcal{A}= [x_0] \cup [x]$), and $x_0$ denotes the 
leftmost site of the chain.
The computation of $\mathcal{I}_x$  involves two-point correlation functions and it quantifies  the  total  amount  of  classical  and  quantum  correlations in the system~\cite{PhysRevB.90.045424,1742-5468-2012-01-P01023,PhysRevB.82.100409,PhysRevB.90.220202,PhysRevLett.111.017201,PhysRevE.62.3096}. The QMI for two sites has been shown to be a useful 
probe to detect a MIT. Particularly, in an exponentially localized phase, $\mathcal{I}_x$ decays exponentially with the distance ($\mathcal{I}_x \sim e^{-x/\xi_1}$). Instead, in the presence 
of an algebraic localization, we expect $\mathcal{I}_x$ to have a power-law behavior ($\sim 1/x^{\beta_1}$). Moreover, for $V=0$, $\beta_1=\alpha$ since the behavior of the QMI is dominated by the decay of the single-particle wavefunctions~\cite{Giu17}
($\psi_E (x)\sim 1/x^\alpha$).
\begin{figure}
	\includegraphics[width=1.\columnwidth]{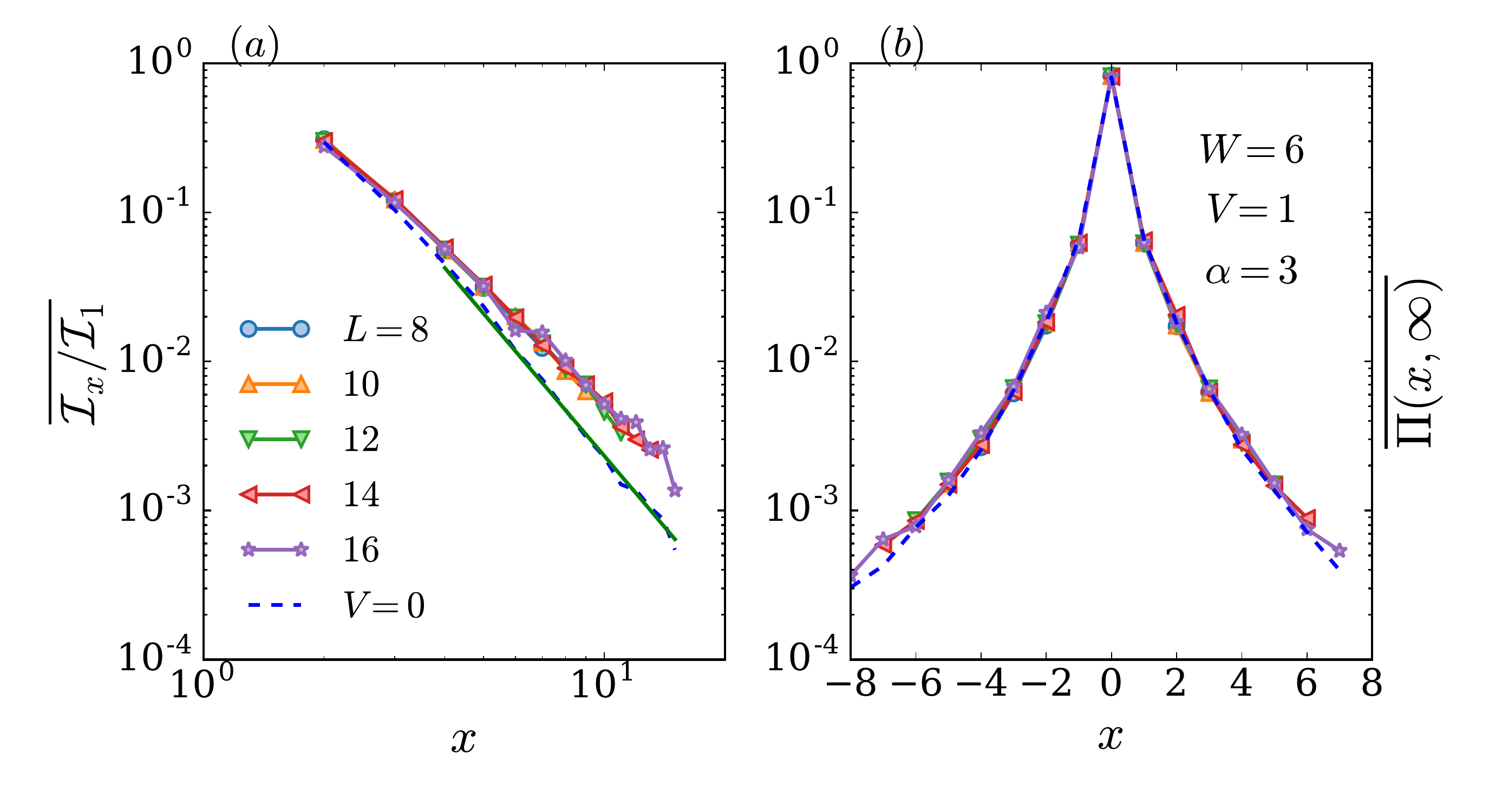}
	\caption{(a): QMI as a function of the distance $x$ for several system sizes $L$, $\alpha=3$ and $W=6$. The dashed line represents the non-interacting case ($\mathcal{I}_x\sim 1/x^{\alpha}$) for $L=16$. 
	(b): Long-time limit of the density-density correlator also for $W=6$ and $\alpha=3$. 
	The dashed line is $\overline{\Pi(x,\infty)}$ for $V=0$ and $L=16$.} 
	\label{fig:Fig3}
\end{figure}

We benchmark the behavior of the QMI for the power-law random banded matrix model ($V=0$), as shown in Fig.~\ref{fig:Fig3}~(a) (dashed line).
For $V=0$ (dashed line), as the system is algebraically localized, $\mathcal{I}_x$ decays algebraically with the distance $x$ ($\sim 1/x^{\beta_1}$), making the QMI not just a useful probe to distinguish between extended and localized states but also 
to inspect different kinds of localized phases. Moreover, the decay rate $\beta_1(V=0)\approx \alpha$, in good agreement with our prediction. 

Figure~\ref{fig:Fig3}~(a) shows the QMI in the localized phase for the interacting case of $\hat{H}$ ($V=1$) for several system sizes $L$. 
Here, also the QMI decays algebraically with the distance $x$ (Fig.~\ref{fig:Fig3}~(a)), and with a rate smaller than one of the non-interacting case ($\beta_1(V=0)=\alpha < \beta_1(V\ne 0)$) as expected. 
The Pinsker's inequality~\cite{1742-5468-2012-01-P01023,PhysRevB.92.180202} gives an upper bound for correlation functions in terms of the QMI, implying that in our case all the two-point correlation functions decay 
at least as fast as $\mathcal{I}_x$. 
Nevertheless, it is natural to expect that generic two-point correlation functions will decay also algebraically with the distance. 

While the existence of an adiabatic connection between the non-interacting and the interacting LIOMs in the localized phase is established~\cite{Ros15,Imbrie2016}, an efficient construction of the latter remains 
still a challenging task. 
Nevertheless, we can get insights into $\{\hat{\tau}_x\}$ by 
studying the long-time limit of the time-dependent density-density correlator~\cite{Proz08,Karr17,Reich15,0953-8984-6-35-001}, defined as
\begin{equation}
 \Phi (x,t) = \frac{1}{\mathcal{N}}\text{Tr}\left [ (\hat n_x(t)-\frac{1}{2} )(    \hat n_0 -\frac{1}{2} )\right ],
\end{equation}
which carries information about the spread of correlation in the system. Moreover, its long-time limit
\begin{equation}
 \Phi (x,\infty) = \lim_{T\rightarrow \infty} \frac{1}{T} \int_0^T \Phi (x,t) dt,
\end{equation}
gives direct access to the integrals of motion as shown in Ref.~\onlinecite{Chandran15}. The time-averaged operator 
\begin{equation}
\hat{n}_x^{\text{t.-av.}}=   \lim_{T\rightarrow \infty} \frac{1}{T}\int_0^T \hat{n}_x(t) dt,
\end{equation}
is an integral of motion ($[\hat{H} , \hat{n}_x^{\text{t.-av.}}]=0$), and thus $\Phi (x,\infty) = \frac{1}{\mathcal{N}}\text{Tr}\left [ (\hat n_x^{\text{t.av.}}-\frac{1}{2} )(    \hat n_0 -\frac{1}{2} )\right ]$ (Eq.~\ref{spred_int}).
\begin{figure*}
	\includegraphics[width=1.\textwidth]{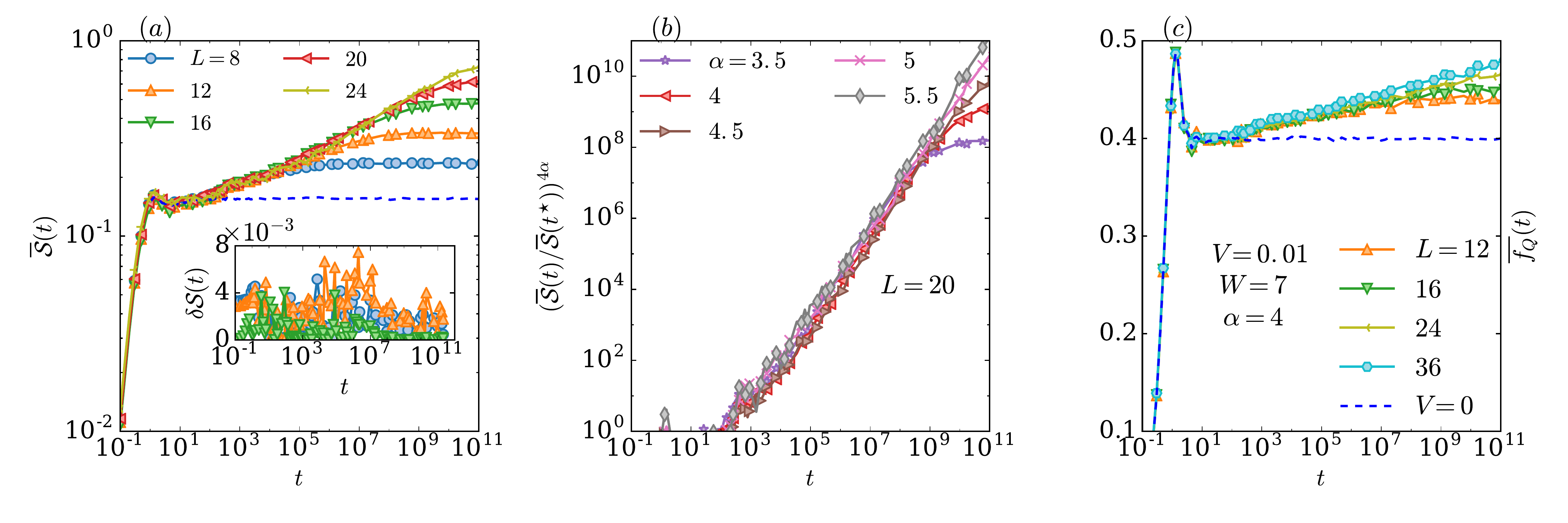}
	\caption{(a): Disorder averaged bipartite entanglement entropy $\overline{\mathcal{S}}(t)$ after a quantum quench starting from $|\psi\rangle = \prod_{s}^{L/2} \hat{c}^{\dagger}_{2s} |0\rangle$ for several system sizes $L$, $\alpha = 4$, $W=7$ and 
	weak interactions $V=0.01$.
	The dashed line is $\overline{\mathcal{S}}(t)$ for the non-interacting case ($V=0$) for $L=24$.  (b): $\overline{\mathcal{S}}(t)$ for fixed $V=0.01$, $W=7$, $L=20$ for several $\alpha$. $\overline{\mathcal{S}}(t)$ 
	has been resclated to underline that in the weak interaction regime 
	$\overline{\mathcal{S}}(t)\sim t^{1/4\alpha}$. (c): Quantum Fisher information $\overline{f}_Q(t)\sim \log{t}$ for several $L$, $\alpha=4$, $W=7$ and $V=0.01$. 
	The dashed line is $\overline{f}_Q(t)$ for the non-interacting case ($V=0$) and $L=36$. 
	All panels are obtained using $\hat{H}^{\text{eff}}$.} 
	\label{fig:Fig4}
\end{figure*}

We renormalize the correlator $\Phi(x,t)$  via its Fourier transform $\Phi(q,t) = \mathcal{F}_q[\Phi(x,t)]$~\footnote{The discrete Fourier transform of a function $f(x)$ is defined by: $\mathcal{F}_q[f(x)] =f(q) = \sum_{x=0}^{L-1} f(x) e^{-iqx}$ with $q = \frac{2\pi}{L}j$, and $0\le j \le L-1$.}
\begin{equation}
 \Pi(x,t) = \mathcal{F}_x \left[ \frac{\Phi(q,t)}{\Phi(q,t=0^+)} \right ],
\end{equation}
after imposing $\lim_{q\rightarrow 0} \frac{\Phi(q,t)}{\Phi(q,t=0^+)} = 1$~\cite{Bera17}. In this way, $\Pi(x,t)$ has a clear interpretation in terms of the wave-packet spreading, since $\lim_{t\rightarrow 0^+}  \Pi(x,t) = \delta_{x,0}$, $\sum_x\Pi(x,t)=1$, and  $\Pi(x,t)\ge 0$~\cite{Bera17}. 

Figure~\ref{fig:Fig3}~(b) shows the long-time limit of $\Pi(x,t)$ in a linear-logarithmic scale as a function of the distance $x$ for a fixed disorder strength $W=6$ in the localized phase of $\hat{H}$, underlining that the decay is 
slower than an exponential. Indeed, for $V=0$ (dashed line) the decay of $\Pi(x,\infty)$ is consistent with a power-law decay ($\Pi(x,\infty)\sim 1/|x|^{\beta}$), as expected, since the single-particle wavefunctions are algebraically localized. 
For the interacting case ($V\ne0$), $\Pi(x,\infty)$  has also a decay that is slower than exponential and it could be algebraic, 
agreeing with the algebraic decay of two-point correlations functions.  It is important to mention, that due to the limitation in system size of exact diagonalization techniques, we are not 
able to reliably extract the decay-rate of $\Pi(x,\infty)$. 
Furthermore, it is worth noting that the difference of $\Pi(x,\infty)$ between the interacting case and the non-interacting one is relatively small and it is visible only for large $x$, even if the interaction strength $V$ is $1/6$ of the disorder strength $W$. 
The last remark could indicate that the adiabatic connection between the non-interacting integrals of motion and the interacting ones could still be valid also in the algebraically localized phase, implying that the $\{\hat{\tau}_x(V\ne 0)\}$ could be constructed 
perturbativly starting from $\{\hat{\tau}_x(V=0)\}$. 

\section{Information propagation}
\label{sec:sec_4}
We now show the implications of algebraic MBL on quantum information propagation.
We restrict our study to the regime of strong disorder and weak interactions, so that we can apply a recent method developed in Ref.~\onlinecite{Giu18}. 
This method is based on perturbation theory and gives an efficient solution of the dynamics of weakly interacting localized systems, pinning down the essential mechanisms for the information propagation in an MBL-phase.

As a first approximation, in the limit of weak
interactions, one could take the integrals of motion to be
the ones of the non-interacting model $\{\hat{\tau}_x = \hat{\eta}_x^\dagger \hat{\eta}_x \}$, thus obtaining the effective Hamiltonian
\begin{equation} 
 \hat{H}^{\text{eff}} = \sum_x \epsilon_x \hat{\eta}^\dagger_x \hat{\eta}_x + V \sum_{x,y} \mathcal{B}_{x,y} \hat{\eta}^\dagger_x \hat{\eta}_x \hat{\eta}^\dagger_y \hat{\eta}_y,
 \label{eq:effective}
\end{equation}
where $\{\hat{\eta}_x^\dagger\}$ ($\{\hat{\eta}_x\}$) defined in Sec.~\ref{sec:sec_2}, are the fermionic creation (annihilation) operators of the single-particle eigenstates $\{\psi_x\}$ of $\hat{H}(V=0)$ 
with single-particle energy $\epsilon_x$. The coefficients $\{\mathcal{B}_{x,y}\}$ can be found perturbatively in the limit of weak interactions~\cite{Giu18}
\begin{equation}
\begin{split}
 \mathcal{B}_{x,y} = \sum_{z} [& |\psi_x(z)|^2  |\psi_y(z+1)|^2  \\ &- \psi_x(z) \psi_y (z) \psi_x (z+1) \psi_y (z+1) ].
 \end{split}
\end{equation}
Importantly, the model $\hat{H}^{\text{eff}}$ (Eq.~\ref{eq:effective}) in this representation is a classical Ising model, and its eigenstates 
are the non-interacting Slater determinant states. 
The role of the term $\sum_{x,y} \mathcal{B}_{x,y} \hat{\eta}^\dagger_x \hat{\eta}_x \hat{\eta}^\dagger_y \hat{\eta}_y$ is to weakly correlate
the eigenenergies of the non-interacting model, reproducing the dephasing mechanism of an interacting localized phase, which is responsible for the propagation of entanglement. 
Moreover, since  the single-particle eigenstates decay in space with the same rate of the hopping constants ($\psi_x(z)\sim 1/|x-z|^{\alpha}$), 
we have $\mathcal{B}_{x,y} \sim 1/|x-y|^{4 \alpha}$ as first approximation. 

The Heisenberg equation for the creation operator $\hat{\eta}_x^\dagger$ reads
\begin{equation}
\label{Hei_eq_eff}
\begin{split}
  \frac{d\hat{\eta}_x^\dagger}{dt} &= i[\hat{H}^{\text{eff}},  \hat{\eta}_x^\dagger]\\
&= i\epsilon_x \hat{\eta}^\dagger_x + i V\sum_y \tilde{\mathcal{B}}_{x,y} \hat{\eta}_y^\dagger \hat{\eta}_y \hat{\eta}^\dagger_x,
  \end{split}
\end{equation}
where $\tilde{\mathcal{B}}_{x,y} = \mathcal{B}_{x,y} +  \mathcal{B}_{y,x}$.
The solution of Eq.~\eqref{Hei_eq_eff} is given by
\begin{equation}
 \hat{\eta}_x^\dagger(t) = e^{+i t\epsilon_x  +itV \sum_y \tilde{\mathcal{B}}_{x,y} \hat{\eta}^\dagger_y \hat{\eta}_y} \hat{\eta}^\dagger_x.
\end{equation}   
The knowledge of the time dependence of the operators $\{\hat{\eta}_x^\dagger\}$ and the use of the Wick's theorem to calculate scalar products between many-body states
allow us to inspect larger system size than the ones accessible by the exact diagonalization at arbitrary time scales.

We first inspect the bipartite entanglement entropy, $\mathcal{S}(t)$, after a global quench starting from $|\psi \rangle = \sum_{s}^{L/2} \hat{c}^\dagger_{2x}|0\rangle$~\cite{Bar12,Aba13}, as a dynamical probe for information propagation.
In Fig.~\ref{fig:Fig4}~(a)~(inset) we compare the exact dynamics computed using $\hat{H}$ (Eq.~\ref{eq:Ham}) with the approximate one using the effective Hamiltonian $\hat{H}^{\text{eff}}$ (Eq.~\ref{eq:effective}). 
We calculate the relative error $\delta \mathcal{S}(t) = \overline{|\mathcal{S}(t)-\mathcal{S}^{\text{approx}}(t)|}/\overline{\mathcal{S}}(t)$, where 
$\mathcal{S}(t)$ and $\mathcal{S}^{\text{approx}}(t)$ are the entanglement entropies computed using $\hat{H}$ and $\hat{H}^{\text{eff}}$ respectively. 
In the weak interaction regime ($V=0.01$) at strong disorder, $W=7$, and $\alpha=4$, $\delta \mathcal{S}(t)$ does not increase with $L$ or with $t$, and it is below $1\%$, thus giving evidence that 
at least for these parameters, our approximation (Eq.~\ref{eq:effective}) is reliable. 
In what follows, we focus our study on the weakly interacting regime at strong disorder, 
thus benefiting from the efficiency of the dynamics simulation generated with $\hat{H}^{\text{eff}}$.
\begin{table*}[t]
\label{ta:fin}
 \begin{tabular}{ l | c | c | c | c| r }
  & $\qquad\text{LIOMs}\qquad$ & $\qquad \mathcal{I}_x \qquad $ & $\quad I(t)\qquad $ & $\qquad \mathcal{S}(t)\qquad $ & $\qquad f_Q(t)\qquad $\\
  \hline			
   $\quad$ & $\quad$ & $\quad$ & $\quad$ & $\quad$ & $\quad$ \\
 \textit{algebraic} MBL & $\text{Tr}[\hat{\tau}_x \hat{A}_y] \sim 1/|x-y|^{\beta}$ & $\sim 1/x^{\beta_1}$ & $I(t)>0$ & $\sim t^{\gamma}$ & $\sim \log{t}$ \\
 $\quad$ & $\quad$ & $\quad$ & $\quad$ & $\quad$ \\
  \textit{exponential} MBL & $\text{Tr}[\hat{\tau}_x \hat{A}_y] \sim e^{-|x-y|/\xi}$ & $\sim e^{-x/\xi_1} $ & $I(t)>0$ & $\sim \log{t}$ & $\sim \log{\log{t}}$ \\
   $\quad$ & $\quad$ & $\quad$ & $\quad$ & $\quad$ \\
  \hline
 \end{tabular}
 \caption{Summary of the main differences between an algebraic and an exponential MBL-phase ($\alpha\rightarrow\infty$).}
\end{table*}

Figure~\ref{fig:Fig4}~(a) shows the disorder averaged entanglement entropy $\mathcal{S}(t)$ for several system sizes $L$ at strong disorder $W=7$ and weak interactions $V=0.01$. The results are obtained by evolving 
the initial state $|\psi\rangle$ with the effective Hamiltonian $\hat{H}^{\text{eff}}$, 
allowing us to present converged curves over many  decades in  time ($t\lesssim 10^{8}$) up to $L=24$. $\overline{\mathcal{S}}(t)$ has an algebraic propagation in time  ($\overline{\mathcal{S}}(t)\sim t^{\gamma}$), and 
its long-time limit has a volume-law scaling, $\lim_{t\rightarrow \infty} \overline{\mathcal{S}}(t)\sim L$ (but not thermal). In the non-interacting case $V=0$, $\overline{\mathcal{S}}(t)$ saturates in the long-time limit to an $L$ independent value, showing that the algebraic growth is only due to 
interactions at least for $\alpha\ge 3$~\cite{PhysRevB.95.094205}. The origin of this algebraic spread of information is due to a dephasing mechanism induced by the interactions which are able to entangle degrees of freedom far away in space, as explained in Ref.~\onlinecite{Pep13}.
The dephasing time at which two localized degrees of freedom at distance $x$ get correlated is given by $t_{\text{deph}}\sim \mathcal{B}_{0,x}^{-1} \sim |x|^{4\alpha}$, leading to an algebraic growth of the 
entanglement entropy $\mathcal{S}(t)\sim t^{1/4\alpha}$ ($\gamma \sim 1/4\alpha$). 
Figure~\ref{fig:Fig4} (b) shows $\overline{\mathcal{S}}(t)$ for several $\alpha$ and fixed system size $L=20$. 
For all inspected values of $\alpha$, the entanglement entropy grows algebraically with time ($\mathcal{S}(t)\sim t^{\gamma(\alpha)}$). 
Moreover, the curves in Fig.~\ref{fig:Fig4} (b) have been rescaled to show that $\gamma \sim 1/{4\alpha}$, obtaining an $\alpha$ independent error of $\approx 15 \% $, this discrepancy 
could be due to the too crude approximation $\mathcal{B}_{x,y}\sim 1/|x-y|^{4\alpha}$. 

Furthermore, we investigate the transport of information via the quantum Fisher information (QFI) for the charge-density wave 
\begin{equation}
\label{eq:QFI}
 f_Q(t) = \frac{4}{L} \left[ \langle \hat{\mathcal{O}}(t)^2 \rangle - \langle \hat{\mathcal{O}}(t) \rangle^2 \right], \quad \hat{\mathcal{O}} = \sum_x (-1)^x \hat n_x .
\end{equation}
The QFI is an experimentally accessible measure, involving only two-point functions, and it quantifies the spread of correlations, measuring quantum-fluctuations of an operator $\hat{\mathcal{O}}$~\cite{Hauke2016,Smith2016, Petz,PhysRevLett.72.3439, 2014Strobel, 2012Toth, 2012Hyllus}.  
Recently, it has been shown that in an exponential MBL-phase (i.e. $\alpha\rightarrow \infty$), $f_Q(t)$ could have a sub-logaritimically slow propagation ($f_Q(t) \sim \log{\log{t}}$), while 
in an Anderson localized phase it saturates quickly with time ($\lim_{t\rightarrow \infty} f_Q(t)\sim \mathcal{O}(L^0)$). Thus, the QFI can be used to distinguish a non-interacting localized phase from an 
interacting one. We now present the results for the QFI propagation in the algebraically localized phase of $\hat{H}$. 
Figure~\ref{fig:Fig4}~(c) shows the disorder averaged $f_Q(t)$ for the non-interacting and weakly interacting case for several system sizes up to $L=36$, by evolving the state $|\psi\rangle$ with $\hat{H}^{\text{eff}}$. 
For the non-interacting case at $\alpha=4$, $\overline{f_Q}(t)$ saturates with time (Fig.~\ref{fig:Fig1}~(c), dashed line), showing a similar behavior like in an Anderson insulator ($V=0$ and $\alpha\rightarrow\infty$)~\cite{Giu18}. 
In the algebraic MBL-phase,  $\overline{f_Q}(t)$ after a quick propagation which overlaps with the non-interacting case, at times $\sim \mathcal{O}(V^{-1})$ interaction effects set in, producing a logaritimic growth of the QFI, 
as shown in Fig.~\ref{fig:Fig4}~(c). This result should be compared to the $\alpha\rightarrow \infty$ limit, in which the propagation of $f_Q(t)$ is parametrically slower ($\sim \log{\log{t}}$). 

In Ref.~\onlinecite{Smith2016} $f_Q(t)$ has been measured experimentally in a long-range interacting disordered quantum Ising model, finding that it grows logarithmically with time. 
The long-range interaction could introduce an algebraic dephasing term, like $\sum_{x,y} \mathcal{B}_{x,y} \hat{\eta}^\dagger_x \hat{\eta}_x \hat{\eta}^\dagger_y \hat{\eta}_y$ in Eq.~\ref{eq:effective} 
with $\mathcal{B}_{x,y}\sim 1/|x-y|^\beta$, which will reproduce 
the logaritimic growth of $f_Q(t)$.

\section{Conclusions}
\label{sec:sec_5}
In this work, we investigated the stability of an MBL-phase in a model possessing algebraically localized single-particle eigenstates. 
In particular, we studied a spinless fermionic Hubbard chain with long-range random hopping and short-range interactions. The non-interacting case is known as power-law random banded matrix model and 
for $\alpha>1$ (power-law decay exponent of the hopping), its single-particle eigenstates are algebraically localized, meaning that they develop long-range power-law tails. 

We provided numerical evidence that an interacting localized phase can still persist once the algebraically localized single-particle states interact via short-range density-density interactions, 
inspecting several MBL markers. At sufficiently strong disorder, the level statistics of energy levels is Poissonian, fluctuation in eigenstates of local observables 
do not scale to zero with system size violating the eigenstate thermalization hypothesis, and finally the bipartite entanglement entropy of eigenstates shows the typical area-law scaling of a localized phase.

Moreover, in order to give further confirmation of the existence of an MBL-phase, we inspected the non-equilibrium dynamical properties of the system, focusing on the relaxation of an initial charge-density wave. 
We observed that no relaxation takes place, meaning that some memory of the local structure of the initial state is preserved over large time scales, which signifies a breakdown of ergodicity.

This MBL-phase can be characterized by the existence of an extensive number of integrals of motion, which develop algebraic decaying tails, thus implying an adiabatic connection with the 
integrals of motion of the non-interacting case. 
In particular, in this algebraic MBL the quantum mutual information between two sites decays as a power-law with the distance, and the long-time limit of the density-density correlator, which provides a direct access to the integrals of motion, 
has algebraically decaying tails.

We then focus on the characterization of information propagation in the algebraic MBL-phase. In this phase, interactions couple the power-law localized integrals of motion, producing a dephasing mechanism which  
generates an unbounded algebraic growth of entanglement entropy. In the limit of weak interactions we relate the rate of the growth with the power of the algebraic decay of the single-particle states. 
We also probed quantum information transport through the propagation of the quantum Fisher information after a quantum quench, showing that it grows logaritimically slowly in time. 
We commented the last result in view of a recent experiment in trapped ions setup, where the dynamics of the quantum Fisher information for a long-range interacting Ising was measured. 

Finally, table~\ref{ta:fin} summarizes the main differences between an MBL-phase with exponentially localized single-particle states (\textit{exponential} MBL-phase) and the algebraically localized one. 
In both phases ergodicity breaks down, and even though particle transport is absent, information can still propagate through the system. Nevertheless, the algebraically localized single-particle states change the 
interaction-induced dephasing mechanism, allowing a faster information propagation than in an exponential MBL-phase. 

\section{Note}
In completing the manuscript we have become aware of a related work on algebraic MBL~\cite{2018arXiv181009779B}. 

In Ref.~\onlinecite{PhysRevB.95.094205} a similar model with deterministic hopping ($J_{x,y} = J/|x-y|^{\alpha}$) has been considered. Its non-interacting limit, called the Burin-Maksimov model~\cite{2018arXiv181001492N}, hosts algebraically localized single-particle states for any $\alpha$ and it has also a zero-density 
of delocalized states~\cite{PhysRevLett.120.110602,2018arXiv181001492N}. Inspecting system sizes up to $L=16$ at strong disorder $W\approx 20$ in Ref.~\onlinecite{PhysRevB.95.094205} (e.g. Fig.~3~(k)) it was concluded that the entanglement entropy grows logarithmically slowly with time ($\mathcal{S}(t)\sim \log{t}$). 
We believe, that for the chosen parameters their system sizes are too small to resolve the algebraic growth of entanglement. Moreover, the role of algebraically localized single-particle states was not discussed.  

\section{Acknowledgments}
We thank J.H.~Bardarson, S.~Bera, H.~Burau, A.~Burin, M.~Heyl, I.M.~Khaymovich, T.L.M.~Lezama, F.~Pollmann, S.N.S.~Prasanna, S.~Roy, D.~Trapin for several illuminating discussions. 

We also express our gratitude to I.M.~Khaymovich, T.L.M.~Lezama, F.~Pollmann, S.~Roy and D.~Trapin for a critical reading of the manuscript. 

\bibliography{Mean_bib}

\end{document}